\documentclass[12pt]{article}%
\usepackage{amsmath,latexsym}
\usepackage{graphicx}
\usepackage{amsmath}
\usepackage{amsfonts}
\usepackage{amssymb}%
\begin{document}

\title{{Macroscopic traversable wormholes with zero tidal
    forces inspired by noncommutative geometry}}
   \author{
Peter K.F. Kuhfittig*\\  \footnote{kuhfitti@msoe.edu}
 \small Department of Mathematics, Milwaukee School of
Engineering,\\
\small Milwaukee, Wisconsin 53202-3109, USA}

\date{}
 \maketitle

\begin{abstract}\noindent
This paper addresses the following issues:
(1) the possible existence of macroscopic
traversable wormholes, given a
noncommutative-geometry background, and
(2) the possibility of allowing zero tidal
forces, given a known density.  It is shown
that whenever the energy density describes
a classical wormhole, the resulting solution
is incompatible with quantum field theory.
If the energy density originates from
noncommutative geometry, then zero tidal
forces are allowed.  Also attributable to
the noncommutative geometry is the
violation of the null energy condition.
The wormhole geometry satisfies the usual
requirements, including asymptotic
flatness.\\

\noindent
PAC numbers: 04.20.-q, 04.20.Jb
\end{abstract}

\section{Introduction}\label{E:introduction}

Wormholes are tunnel-like structures in
spacetime that link widely separated
regions of our Universe or different universes
altogether.  Morris and Thorne \cite{MT88}
proposed the following line element for the
wormhole spacetime:
\begin{equation}\label{E:line1}
ds^{2}=-e^{2\Phi(r)}dt^{2}+\frac{dr^2}{1-b(r)/r}
+r^{2}(d\theta^{2}+\text{sin}^{2}\theta\,
d\phi^{2}),
\end{equation}
using units in which $c=G=1$.  In this line
element, $b=b(r)$ is called the \emph{shape
function} and $\Phi=\Phi(r)$ is called the
\emph{redshift function}, which must be
everywhere finite to avoid an event horizon.
For the shape function we must have
$b(r_0)=r_0$, where $r=r_0$ is the radius of
the \emph{throat} of the wormhole.  A key
requirement is the \emph{flare-out condition}
at the throat: $b'(r_0)<1$, while $b(r)<r$
near the throat.  The flare-out condition can
only be satisfied by violating the null energy
condition, to be discussed later.  Ordinarily,
this violation implies that a wormhole can
only be held open by the use of ``exotic
matter."

Using an orthonormal frame, the Einstein field
equations $G_{\hat{\mu}\hat{\nu}}=8\pi T_
{\hat{\mu}\hat{\nu}}$ yield the following simple
interpretation for the components of the
stress-energy tensor:
$T_{\hat{t}\hat{t}}=\rho(r)$, the energy density,
$T_{\hat{r}\hat{r}}=p_r$, the radial pressure,
and $T_{\hat{\theta}\hat{\theta}}=
T_{\hat{\phi}\hat{\phi}}=p_t$, the lateral
pressure.  Morris and Thorne then suggested the
theoretical construction of a wormhole by using
the following strategy: specify the functions
$b=b(r)$ and $\Phi=\Phi(r)$ in order to produce
the desired properties.  While this strategy
retains complete control over the geometry, it
leads to a practical problem: the engineering
team must be able to come up with the materials
or fields that yield the required stress-energy
tensor.  In this paper we will therefore adopt
a mixed strategy consisting of a combination
of geometric and physical specifications.

The success of the mixed strategy is going to
depend on an important outcome of string theory,
the realization that coordinates may become
noncommutative operators on a $D$-brane
\cite{eW96, SW99}.  Noncommutativity replaces
point-like objects by smeared objects
\cite{SS03, NSS06, mR11} with the aim of
eliminating the divergences that normally
appear in general relativity.  (The
noncommutative geometry results in a fundamental
discretization of spacetime due to the
commutator $[\textbf{x}^{\mu},\textbf{x}^{\nu}]
=i\theta^{\mu\nu}$, where $\theta^{\mu\nu}$ is
an antisymmetric matrix.)

The smearing can be modeled using a Gaussian
distribution of minimal length $\sqrt{\theta}$
instead of the Dirac delta function \cite{NSS06,
mR11, RKRI12, pK13}.  An equally effective way is to
assume that the energy density of the static
and spherically symmetric and particle-like
gravitational source has the form
\begin{equation}\label{E:rho1}
  \rho(r)=\frac{M\sqrt{\theta}}
     {\pi^2(r^2+\theta)^2};
\end{equation}
(see Refs. \cite{LL12} and \cite{NM08}.)
Here the mass $M$ is diffused throughout the
region of linear dimension $\sqrt{\theta}$ due
to the uncertainty.  The noncommutative
geometry is an intrinsic property of spacetime
and does not depend on particular features
such as curvature.

As discussed in Ref. \cite{pKnew}, to describe
the mixed strategy for wormhole construction,
we must first return to the Einstein field
equations $G_{\hat{\mu}\hat{\nu}}=
8\pi T_{\hat{\mu}\hat{\nu}}$, resulting in the
following forms:
\begin{equation}\label{E:Einstein1}
  \rho(r)=\frac{b'}{8\pi r^2},
\end{equation}
\begin{equation}\label{E:Einstein2}
   p_r(r)=\frac{1}{8\pi}\left[-\frac{b}{r^3}+
   2\left(1-\frac{b}{r}\right)\frac{\Phi'}{r}
   \right],
\end{equation}
\begin{equation}\label{E:Einstein3}
   p_t(r)=\frac{1}{8\pi}\left(1-\frac{b}{r}\right)
   \left[\Phi''-\frac{b'r-b}{2r(r-b)}\Phi'
   +(\Phi')^2+\frac{\Phi'}{r}-
   \frac{b'r-b}{2r^2(r-b)}\right].
\end{equation}
Since Eq. (\ref{E:Einstein3}) can be obtained
from the conservation of the stress-energy tensor
$T^{\mu\nu}_{\phantom{\mu\nu};\nu}=0$, only two
of Eqs. (\ref{E:Einstein1})-(\ref{E:Einstein3})
are independent.  As a result, these can be
written in the following form:
\begin{equation}\label{E:E1}
  b'=8\pi\rho r^2,
\end{equation}
and
\begin{equation}\label{E:E2}
  \Phi'=\frac{8\pi p_rr^3+b}{2r(r-b)}.
\end{equation}
The discussion in Ref. \cite{pKnew} assumes
an equation of state, $p_r=\omega\rho$.
Moreover, if $\rho(r)$ is known, an
example of which is Eq. (\ref{E:rho1}),
then $b(r)$ can be determined from Eq.
(\ref{E:E1}).  Unfortunately, since
$b(r_0)=r_0$, we see from Eq. (\ref{E:E2})
that $\Phi'$ (and hence $\Phi$) are not
likely to exist, thereby leading to an
event horizon.  (Exceptions occur for
certain special forms of $b(r)$.)  While
assigning the redshift function $\Phi$
avoids this problem, we can see from
Eqs. (\ref{E:Einstein2}) and
(\ref{E:Einstein3}) that we have returned
to the engineering problem of having to
determine the components
$T_{\hat{r}\hat{r}}$ and
$T_{\hat{\theta}\hat{\theta}}$ of the
stress-energy tensor.

One of the goals in this paper is to show
that macroscopic traversable wormholes
with zero tidal forces may exist given a
noncommutative-geometry background.  Purely
mathematically speaking, this goal can be
accomplished by simply letting $\Phi\equiv
\text{constant}$, so that $\Phi'\equiv 0$,
the \emph{zero-tidal-force solution}
\cite{MT88}.  That is the topic of the
next section.  The feasibility of this
approach will then be discussed in
Sec. \ref{S:feasibility}.

\section{The solution}\label{S:solution}

Our discussion begins with Eq. (\ref{E:rho1}),
\begin{equation}\label{E:rho2}
  \rho(r)=\frac{M\sqrt{\theta}}
     {\pi^2(r^2+\theta)^2},
\end{equation}
which immediately yields the total mass-energy
$M_{\theta}$ of a sphere of radius $r$:
\begin{equation}\label{E:mass}
   M_{\theta}=\int^r_0\rho(r')4\pi (r')^2
   dr'=\frac{2M}{\pi}\left(\text{tan}^{-1}
   \frac{r}{\sqrt{\theta}}-
   \frac{r\sqrt{\theta}}{r^2+\theta}\right).
\end{equation}
Here the very small linear dimension
$\sqrt{\theta}$ raises a question regarding
the scale.  From Eq. (\ref{E:Einstein1}),
\begin{equation}\label{E:shape1}
   b(r)=\frac{8M\sqrt{\theta}}{\pi}
   \int^r_{r_0}\frac{(r')^2dr'}{[(r')^2+\theta]^2}
   +r_0,
\end{equation}
ensuring that $b(r_0)=r_0$.  Since Eq.
(\ref{E:shape1}) is valid for any $r_0$, the
wormhole can be macroscopic, but, as we will
see in the next section, we also prefer a
moderate throat size.  So we simply assume
that $r=r_0$ is just large enough to permit
passage.  The resulting shape function is
\begin{equation}\label{E:shape2}
  b(r)=\frac{4M\sqrt{\theta}}{\pi}
  \left(\frac{1}{\sqrt{\theta}}\text{tan}^{-1}
  \frac{r}{\sqrt{\theta}}-\frac{r}{r^2+\theta}-
  \frac{1}{\sqrt{\theta}}\text{tan}^{-1}
  \frac{r_0}{\sqrt{\theta}}+\frac{r_0}{r_0^2
  +\theta}\right)+r_0.
\end{equation}
Since $\Phi'\equiv 0$, $p_r(r)$ and $p_t(r)$ can
now be obtained directly from Eqs.
(\ref{E:Einstein2}) and (\ref{E:Einstein3}).

\emph{Remark:} Returning to the question of
length scale, it is noted by Nicolini et al.
\cite{NSS06} that there is no need to change
the Einstein tensor in the field equations
since the noncommutative effects can be
implemented by modifying only the stress-energy
tensor, as performed above.  So the length
scale need not be restricted to the Planck
scale.

Next, to check the flare-out condition, we need
to examine
\begin{equation}\label{E:bprime}
  b'(r)=\frac{4M}{\pi}\frac{\sqrt{\theta}}
  {r^2+\theta}+\frac{4M\sqrt{\theta}}{\pi}
  \frac{r^2-\theta}{(r^2+\theta)^2}.
\end{equation}
(Observe that $b'(r)>0$ since $\theta\ll r$.)
 At or near the throat, $b'(r)<1$ as long as
$\sqrt{\theta}\ll M$.  So the flare-out
condition has been met.  (We also have $b(r)<r$
near the throat.)  It should be noted that
$b'(r)$ is relatively small, so that $b(r)$
is a slowly increasing function.

Closely related to the flare-out condition is
the violation of the null energy condition at
or near the throat.  Indeed,
\begin{equation}
  \rho+p_r=\frac{M\sqrt{\theta}}
  {\pi^2(r^2+\theta)^2}-\frac{1}{8\pi}
  \frac{b(r)}{r^3},
\end{equation}
which is negative at or near the throat
because of the small $\theta$.

Finally, $\text{lim}_{r\rightarrow
\infty}\frac{b(r)}{r}=0$, so that the
wormhole spacetime is asymptotically flat.
All the conditions required for the existence
of a wormhole have thereby been met.

\section{Feasibility}\label{S:feasibility}

A question that is always of interest in
wormhole physics is whether a sufficiently far
advanced civilization could in principle
construct a wormhole.  Based on the earlier
discussion, a relatively straightforward way
would be to start with what appears to be
ordinary matter with a known energy density
$\rho(r)$.  By simply fixing $M$ and
$\theta$, we could view Eq. (\ref{E:rho2})
in this light.  The resulting $b(r)$ in
Eq. (\ref{E:shape2}) would in principle
suffice for the construction of the wormhole.

Unfortunately, having to assign $\Phi(r)$
makes the theoretical construction more
problematical since we have now returned
to the engineering problem mentioned earlier.
On the other hand, the assumption that
$\Phi\equiv \text{constant}$ is not only
physically desirable, it also produces a
simple solution.  The feasibility of this
approach is discussed next.

\subsection{The zero-tidal-force assumption}

Regarding the validity of the zero-tidal-force
assumption, the previous section is of no help.
So we need to reconsider a topic that is
usually neglected, the compatibility of
\emph{classical wormhole theory} with quantum
field theory, taken up in some detail in Refs.
\cite{pK08a, pK08b}.   The compatibility of
charged and thin-shell wormholes with
quantum field theory are discussed in Refs.
\cite{pK11, pK12}, respectively.  In all
cases, a wormhole must satisfy an extended
version of the quantum inequalities,
originally proposed by Ford and Roman
\cite{FR96}, who also showed that the
exotic matter must be confined to a narrow
band around the throat.  In other words,
the highly problematical exotic region
should be made as small as possible.  It
is shown in Refs. \cite{pK08a, pK08b} that
this can only be accomplished by
fine-tuning the metric coefficients.  More
precisely, to satisfy the extended quantum
inequalities, one must strike a balance
between reducing the size of the exotic
region and the degree of fine-tuning of
the metric coefficients required to
achieve this reduction.  In particular,
$\Phi'(r)$ has to be fine-tuned to remain
in a narrow range.  The most important
conclusion for present purposes is that the
value $\Phi'(r)\equiv 0$ is outside this
range, so that the resulting wormhole
cannot be compatible with quantum field
theory.  This finding also explains why
none of the wormhole solutions in Ref.
\cite{MT88}, which assumes
$\Phi'(r)\equiv 0$ throughout, satisfy
the quantum inequalities, first pointed
out in Ref. \cite{FR96}.

\subsection{Noncommutative geometry}

The previous subsection dealt with the
energy density of ordinary matter and
subsequent incompatibility with quantum
field theory.  The situation is entirely
different for $\rho(r)$ in Eq.
(\ref{E:rho1}), being a consequence of
noncommutative geometry: the quantum
inequalities, which assume an inertial
Minkowski spacetime without boundary,
no longer apply.  So the ability to
assume zero tidal forces becomes a
critical consequence of the
noncommutative-geometry background.  An
equally important consequence is that
the violation of the null energy
condition can largely be attributed
to this geometric background rather
than to ``exotic matter."  Thus, since
$\sqrt{\theta}\ll M$,
\begin{equation*}
  \rho(r)+p_r(r)=\frac{M\sqrt{\theta}}
  {\pi^2(r^2+\theta)^2}+\frac{1}{8\pi}
  \left[-\frac{b}{r^3}+2\left(1
  -\frac{b}{r}\right)\frac{\Phi'}{r}
  \right]<0
\end{equation*}
near $r=r_0$ even if $\Phi'(r)\neq0$.
At the very least, the condition would
hold for a large class of redshift
functions.

\subsection{More fine-tuning}

Based on the above results, we conclude that
if noncommutative geometry is a correct model,
then the laws of physics seem to allow
macroscopic traversable wormholes with zero
tidal forces, while classical general
relativity does not.

It becomes apparent, however, that from the
standpoint of physical construction by an
advanced civilization, there are additional
practical considerations reminiscent of the
fine-tuning problems in Refs. \cite{pK08a,
pK08b}: due to the smearing effect,
$b(r)$ in Eq. (\ref{E:shape2}) is a very
slowly increasing function of the radial
coordinate $r$, as a result of which the
total mass inside a sphere of radius $r$
increases very slowly as well.  So the
construction would require considerable
fine-tuning even for relatively small
throat sizes.  More fine-tuning would be
required just to determine $\rho(r)$.
(We may assume that a sufficiently far
advanced civilization could make such a
determination, even if this is well
beyond our own capability.)
Sufficiently close to Einstein gravity,
i.e., for sufficiently small $\theta$,
the construction becomes ever more
difficult, as one would expect.

\section{Zero tidal forces: other motivations}

While our main concern regarding zero
tidal forces is traversability, other
motivations exist.  For example, in the
discussion of self-contained phantom
wormholes in semi-classical gravity,
Garattini and Lobo \cite{GL07} assumed the
equation of state $p_r(r)=\omega(r)\rho(r)$,
$\omega(r)<-1$.  The variable parameter
$\omega(r)$ generalizes the cosmological
equation of state $p=\omega\rho$,
$\omega<-1$, representing phantom energy.
The restricted choices for $\Phi(r)$ and
$\omega(r)$ discussed include the constant
redshift function which results in a
reduced form of the curvature scalar:
$R(r)=2b'(r)/r^2$.  (See Ref. \cite{GL07}
for details.)

In the present paper, the case for zero tidal
forces can also be strengthened if a
noncommutative-geometry background is assumed
in a slightly  modified gravitational
theory, as defined  in Ref. \cite{kuhf}.
The definition is  based on the
gravitational field equations  in the form
used by Lobo and Oliveira  \cite{LO09} for
$f(R)$ modified gravity under the assumption
that $\Phi'(r)\equiv0$:
\begin{equation}\label{E:Lobo}
   \rho(r)=F(r)\frac{b'(r)}{r^2},
\end{equation}
\begin{equation}
   p_r(r)=-F(r)\frac{b(r)}{r^3}
   +F'(r)\frac{rb'(r)-b(r)}{2r^2}
   -F''(r)\left[1-\frac{b(r)}{r}\right],
\end{equation}
\begin{equation}
   p_t(r)=-\frac{F'(r)}{r}\left[1-\frac{b(r)}{r}
   \right]+\frac{F(r)}{2r^3}[b(r)-rb'(r)],
\end{equation}
where $F=\frac{df}{dR}$.  The curvature scalar
$R(r)$ is again given by
\begin{equation}\label{E:Ricci}
   R(r)=\frac{2b'(r)}{r^2}.
\end{equation}
Comparing Eqs. (\ref{E:Lobo}) and
(\ref{E:Ricci}), a slight change in $F$
results in a slight change in $R$,  which
characterizes $f(R)$ modified gravity.  As
noted in Ref. \cite {kuhf}, we may quantify
the notion of slightly modified gravity by
assuming that $F(r)$ remains close to unity
and relatively ``flat," i.e., both $F'(r)$
and $F''(r)$ remain relatively small in
absolute value.

To apply these ideas to the present study,
it is sufficient to examine Eq. (\ref{E:Lobo}):
\begin{equation}
  b'(r)=\frac{r^2}{F(r)}\frac{M\sqrt{\theta}}
  {\pi^2(r^2+\theta)^2}.
\end{equation}
Assuming that $F(r)>0$ for all $r$, we obtain
$0<b'(r)<1$ since $\sqrt{\theta}\ll M$.  So
the shape function satisfies the flare-out
condition.  Since we are now dealing with
$f(R)$ modified gravity, the quantum
inequalities do not apply, once again
rescuing the zero-tidal-force assumption.

\section{Conclusion}

This paper deals with the possible existence
and, in a limited way, the theoretical
construction of macroscopic
traversable wormholes in a
noncommutative-geometry setting.
We relied on a mixed strategy by assuming
a known energy density $\rho(r)$ and then
assigning the redshift function $\Phi(r)$.
(Attempting to determine $\Phi(r)$ from the
Einstein field equations often leads to an
event horizon.)  Once $\Phi(r)$ is assigned,
calculating the pressure leads to the
engineering problems discussed in Ref.
\cite{MT88}: one has to determine the
materials or fields that lead to the
remaining components of the stress-energy
tensor.

One of the goals in this paper is to
obtain a macroscopic wormhole with zero
tidal forces, i.e., $\Phi'(r)\equiv 0$.
If $\rho(r)$ is known, then the
zero-tidal-force assumption allows the
determination of both $p_r$ and $p_t$
directly.  The energy density $\rho(r)$
may correspond to ordinary matter in
classical wormhole theory, but as we
saw in Sec. \ref{S:solution},
we may also have
\[
   \rho(r)=\frac{M\sqrt{\theta}}
   {\pi^2(r^2+\theta)^2},
\]
based on the assumption that in
noncommutative geometry, point-like
particles are replaced by smeared objects,
so that the mass $M$ is diffused throughout
the region due to the uncertainty.

The difference between the two scenarios is
that in the classical case, the
zero-tidal-force assumption makes the
wormhole solution incompatible with quantum
field theory.  More precisely, this solution
cannot satisfy the extended quantum
inequalities.  Since these are based on
an inertial Minkowski spacetime without
boundary, they do not apply to wormholes
in noncommutative geometry.  So the
zero-tidal-force solution is allowed.
It was also determined that, under
fairly general conditions, the
violation of the null energy condition
is due to the noncommutative geometry,
rather than to exotic matter.

As discussed in Sec. \ref{S:feasibility},
for classical wormholes to be compatible
with quantum field theory, one must strike
a balance between reducing the size of the
exotic region and the degree of fine-tuning
required to achieve this reduction.  While
this problem is avoided in noncommutative
geometry, fine-tuning would still be
required in the theoretical  construction
of the wormhole because the shape function
$b(r)$ rises very slowly due to the
smearing effect, even with relatively small
throat sizes.  Even more fine-tuning may be
required to determine $\rho(r)$.

\end{document}